\documentclass{aa}  
\usepackage{graphicx}
\usepackage{txfonts}
\usepackage{amsmath}	
\usepackage{xcolor}
\usepackage{cancel}
\usepackage{multirow}
\usepackage{multicol}

\usepackage{hyperref}

\begin{document}

   \title{Stream Automatic Detection with Convolutional Neural Network}
   \subtitle{(SAD-CNN)}

   \author{Alex Vera-Casanova. \inst{1}
        \and Monsalves Gonzalez N. \inst{1}
        \and Facundo A. Gómez. \inst{1}
         \and Marcelo Jaque Arancibia. \inst{1}
         \and Fontirroig, V. \inst{1}
         \and D. Pallero. \inst{2}
         \and R\"{u}diger Pakmor. \inst{3}
         \and Freeke van de Voort. \inst{4}
         \and Robert J. J. Grand. \inst{5}
         \and Rebekka Bieri. \inst{6}
         \and Federico Marinacci. \inst{7,8}
          }

   \institute{
    \label{inst.ULS}Departamento de Astronom\'ia, Universidad de La Serena, Raúl Bitrán Nº 1305, La Serena, Chile. \\
    \email{alex.vera@userena.cl}
    \and
    \label{inst.UTFSM}Departamento de Física, Universidad Técnica Federico Santa María, Avenida España 1600, 2390123, Valparaíso, Chile.
    \and
    \label{inst.MPA}Max-Planck-Institut f\"{u}r Astrophysik, Karl-Schwarzschild-Str. 1, D-85748, Garching, Germany.
    \and
    \label{inst.Cardiff}Cardiff Hub for Astrophysics Research and Technology, School of Physics and Astronomy, Cardiff University, Queen's Buildings, Cardiff CF24 3AA, UK.
    \and
    \label{inst.Zurich}{Center for Theoretical Astrophysics and Cosmology, Department of Astrophysics, University of Zurich}
    \and
    \label{inst.Liverpool}Astrophysics Research Institute, Liverpool John Moores University, 146 Brownlow Hill, Liverpool L3 5RF, UK.
    \and 
    \label{inst.Bologna}Department of Physics and Astronomy “Augusto Righi,” University of Bologna, via Gobetti 93/2, 40129, Bologna, Italy.
    \and
    \label{inst.INAF}INAF, Astrophysics and Space Science Observatory Bologna, Via P. Gobetti 93/3, 40129 Bologna, Italy
}

   \date{Received Month day, 2025; accepted Month day, 2025}

  \abstract
   {Galactic halos host faint substructures, such as stellar streams and shells, which provide insights into the hierarchical assembly history of galaxies. To date, such features have been identified in external galaxies by visual inspection. However, with the advent of larger and deeper surveys and the associated increase in data volume, this methodology is becoming impractical.}
   %However, this methodology is becoming unfeasible with the advent of larger and deeper surveys.} %Detecting these features is challenging due to their low surface brightness (LSB) and complex morphology for next-generation astronomical surveys.}
   {Here we aim to develop an automated method to detect low surface brightness features in galactic stellar halos. Moreover, we seek to quantify its performance when considering progressively more complex data sets, including different stellar disc orientations and redshifts. } %galaxy redshift
   {We develop the Stream Automatic Detection with Convolutional Neural Networks, SAD-CNN. This tool is trained on mock surface brightness maps obtained from simulations of the Auriga Project. The model incorporates transfer learning, data augmentation and balanced datasets to optimise its detection capabilities at surface brightness limiting magnitudes ranging from 27 to 31 mag arcsec$^{-2}$.}
   {The iterative training approach, coupled with transfer learning, allowed the model to adapt to increasingly challenging datasets, achieving precision
    and recall metrics above $80\%$ in all considered scenarios. The use of a well-balanced training dataset is critical for mitigating biases, ensuring that the CNN accurately distinguishes between galaxies with and without streams.}
%  an Precision of $\sim 90\%$ for edge-on galaxies at $z=0$ and $\sim 89\%$ for inclined galaxies. The model effectively reduces False positives caused by disc structures, such as spiral arms, and demonstrates robust generalization to complex scenarios.}
   {SAD-CNN is a reliable and scalable tool for automating the detection of faint substructures in galactic halos. Its adaptability makes it well-suited for future applications, including the analysis of data from upcoming large astronomical surveys (such as LSST, JWT).}


\maketitle{}

\section{Introduction}

    The current paradigm of galaxy formation postulates that accretion of material from the surrounding environment drives the mass growth of galaxies \citep{Searle_Zinn78, White_Rees78,1991ApJ...379...52W, Johnston96, 2008ApJ...689..936J, 2008MNRAS.385.1365L}. Galaxies are not singular bodies, they are surrounded by vast halos that are home to a variety of faint features, indicative of their intricate assembly history \citep{1999Helmi, 2008A&ARv..15..145H, Cooper2010, 2015ApJ...799..184P}. These substructures, often manifested as streams, shells, or plumes, provide crucial information about the accretion history of a galaxy, as they are the remnants of past accretion processes that feed the hierarchical growth of galaxies \citep{1994Natur.370..194I,1995AJ....110..140H, Johnston96, 1999Majewski,2003ApJ...592L..25H,2009Natur.461...66M,2010Marinez-Delgado}. 
    
    Stellar streams are long, narrow structures composed of stars orbiting galaxies. These streams are usually stretched along the progenitor's trajectory, whose stars are pruned due to the tidal forces exerted by the gravitational field of the galaxy around which they orbit \citep{2008ApJ...689..936J}. These streams result from interactions between galaxies and ongoing accretion processes.  Several studies have attempted to characterise the accretion history of our galaxy, the Milky Way, by analysing substructures in the six-dimensional phase space distribution of local volumes, allowing not only the detection of phase-mixed stellar streams but also to place constraints on the structure, size and orbital histories of the progenitors \citep{1999MNRAS.307..495H, 2003MNRAS.339..834H, 2005MNRAS.359...93M, 2010MNRAS.401.2285G, Belokurov2018, Malhan2018, Helmi2020, 2021ApJ...911..149L}. Recently, advances in observational technology, such as the Gaia astrometric satellite, have greatly increased the number of known streams in the Milky Way \citep{Gaia2018,2021ApJ...914..123I}. This information, combined with the mapping of the outer halo with photometric and spectroscopic surveys \citep{Bonaca2025}, has allowed a comprehensive view of the merging history of our Galaxy for the first time.
    
    The study of these faint substructures in external galaxies presents important challenges, mainly attributed to their detectability \citep{2006ApJ...642L.137B, 2010Marinez-Delgado, 2013ApJ...765...28A, 2018A&A...614A.143M, 2018ApJ...862..114S, Gordon2024}. The detection and characterisation of tidal streams is a highly complex task due to their extremely low surface brightness. Indeed, deep observations with surface brightness limit deeper than 29 mag arcsec$^{-2}$ in the r-band are required \citep{bullock_johnston2005, 2013MNRAS.434.3348C, 2000ApJ...529..886C, 2014A&A...566A..97J, 2016MNRAS.456.1359F, 2018A&A...614A.143M, 2018ApJ...857..144H, 2019A&A...632A.122M, Gordon2024}. Previous studies have started to provide different censuses about tidal streams and faint stellar substructures in external galaxies, which are crucial for understanding their prevalence.  \citet{2013ApJ...765...28A}, using a sample of 1781 galaxies, reported that approximately 26\% of galaxies with stellarmasses exceeding $10^{10.5} M_{\odot}$ exhibited detectable tidal features with a $\mu^{lim} \approx 27.7 $ mag arcsec$^{-2}$. Expanding on these findings, \citet{2017IAUS..321..180D} explored a broader census through the MATLAS survey, in a sample of 360 massive nearby galaxies, revealing a higher fraction of galaxies, reaching 40\%, with observed faint substructures. A more recent study based on the RESOLVE survey \citep{2018ApJ...857..144H}, which reaches an r-band depth of $\sim 27.9$ mag arcsec$^{-2}$ for a sample of 1048 galaxies, identified faint features in $\sim17\%$ of their galaxies. \citet{2018A&A...614A.143M} reported a limiting surface brightness of $\mu^{lim}_r \approx 28.11 $ for their sample of 232 images. In particular \citet{Walmsley2019} detect tidal features in a dataset covering approximately 170 deg$^2$ with a limiting surface brightness of $\mu^{lim}_r \approx 27.1 $. 
    Notably, \citet{2020MNRAS.498.2138B} found that the presence of these faint features correlates with the mass of the host galaxy and is influenced by environmental factors. These detection rates highlight the difficulties in establishing a comprehensive census of faint features, which may be affected by factors such as the depth of the survey, the quality of the data, and the specific techniques used for identification. Moreover, the nature of these structures, which can be distorted or disrupted by various factors, further complicates their study. Thus, advanced techniques, such as stacking of multiple images, pixel-level analysis, and Multi-Gauss Expansion (MGE) model are used for the detection and characterisation of tidal streams \citep{2018ApJ...866..103K,2022A&A...662A.124S,Juan_MiroCarretero2023,2024MNRAS.529..810R}.
    
    % In the numerical counterpart, cosmological simulations,
    In the numerical counterpart, such as the Auriga project \citep{AURIGA} and FIRE simulations \citep{2018MNRAS.480..800H}, have been instrumental in modelling the hierarchical growth of galaxies and their stellar halos. These simulations track the accretion histories of galaxies, allowing researchers to study the origins and properties of stellar streams in a cosmological context. Several works have utilised cosmological simulations to explore stellar halos and their properties, which have provided valuable insights into these complex structures \citep{2008ApJ...689..936J, Cooper2010, Tumlinson2010, 2011ApJ...733L...7H, Tissera2013, Tissera2014, 2015MNRAS.454.3542E, 2017MNRAS.464.2882A, 2019MNRAS.485.2589M, Vera-casanova2022, MartinLSST2022, 2022arXiv220808443V, KadoFong2022, 2023ApJ...949...44S, Jenny2024, Tau2024, 2024arXiv241009144R, Shipp2024}. These theoretical approaches offer diverse insights into the formation and evolution of faint features, providing a basis for understanding the processes that drive the assembly history of galaxies. Recently work has focused on specific aspects of stellar halo formation. For example, \citet{2015MNRAS.454.3542E, Walder2024} investigated how properties of stellar tidal streams can provide properties of the dark halo, while \citet{2017MNRAS.464.2882A} explored the connection between a merger history and the properties of the accreted stellar halo. \citet{Vera-casanova2022} used the Auriga simulations to study stellar halos in Milky Way-mass galaxies, predicting the fraction of stellar streams detectable at different surface brightness limits. \citet{MartinLSST2022} emphasized the importance of including realistic observational conditions in simulations, showing how such factors can significantly impact the detectability of faint features. Similarly, \citet{KadoFong2022} demonstrated how stream morphology and spatial distribution can constrain the assembly history of halos in cosmological simulations.
    
    The next generation of surveys, such as LSST and GAIA, \citep{2019ApJ.LSST...873..111I,2023Gaiadr3} are about to increase the volume of data available to researchers dramatically. This burst of data motivates the development of automated tools, particularly those employing machine learning techniques, to extract and analyse information from these large datasets efficiently. Such tools applied to the forthcoming surveys could allow the detection and characterisation of a large number of stellar streams, enabling a statistical study based on the properties and prominence of these substructures. Accurately quantifying the fraction of galaxies that present a low surface brightness (SB) substructure, as a function of limiting surface brightness magnitude, would not only allow us to constrain individual galaxies' merger histories, but could also provide important constraints on the galaxy merger rate as a function of time \citep{Vera-casanova2022, MartinLSST2022, 2024arXiv240903585M}.

    Convolutional Neural Networks (CNNs) belong to a class of neural networks commonly used for processing array representations, such as 1D sequences, 2D images, or 3D videos \citep{2015Natur.521..436L}. In the context of images, they have been widely applied to tasks such as recognition \citep{2021MNRAS.504..372B}, segmentation \citep{Farias2020}, and style transfer \citep{2016arXiv161107865G}. The main idea behind CNNs is to automatically extract relevant features from images and condense this information into a feature map. Then, the model learns the non-linear combinations of these features through a training process, enabling predictions for new input data.
    In recent years, CNNs have become powerful tools for automated feature extraction and pattern recognition in astronomy. They are applied to tasks such as galaxy morphology classification \citep{Walmsley2019, Farias2020, 2021AJ....162..206S, 2021MNRAS.504..372B} and the identification of faint features \citep{Baxter2021, Gordon2024, Fontirroig2024}. By exploiting the enormous amount of data produced by modern astronomical surveys, CNNs provide an efficient and powerful approach to analysing and interpreting astronomical datasets. Recently, \citet{Gordon2024} advanced this field by employing CNNs to classify tidal features into distinct categories, highlighting the potential of automated methods for analysing faint substructures.
    
    In this study, we use the capabilities of CNNs to automate the detection of faint substructures, specifically stellar streams, in galactic halos. By training a CNN on simulated surface brightness maps generated from the Auriga Project simulation dataset, we aim to develop a robust and efficient methodology to identify and characterise stellar streams. Our approach accelerates the process of analysing complex astronomical datasets and allows the exploration of intricate structures within galactic halos.
    The paper is organised as follows. In Section 2, we introduce the simulation and the method employed to make the training samples. Additionally, we describe the CNN utilized in this process. In Section 3 we explain the methodology for training step by step. In section 4 we discuss the training process, while in section 5 we present the results obtained from random samples of images. Finally, in section 6 we summarise the main results obtained in this work.
    
\begin{figure}
    \centering
    \includegraphics[width=0.45\textwidth]{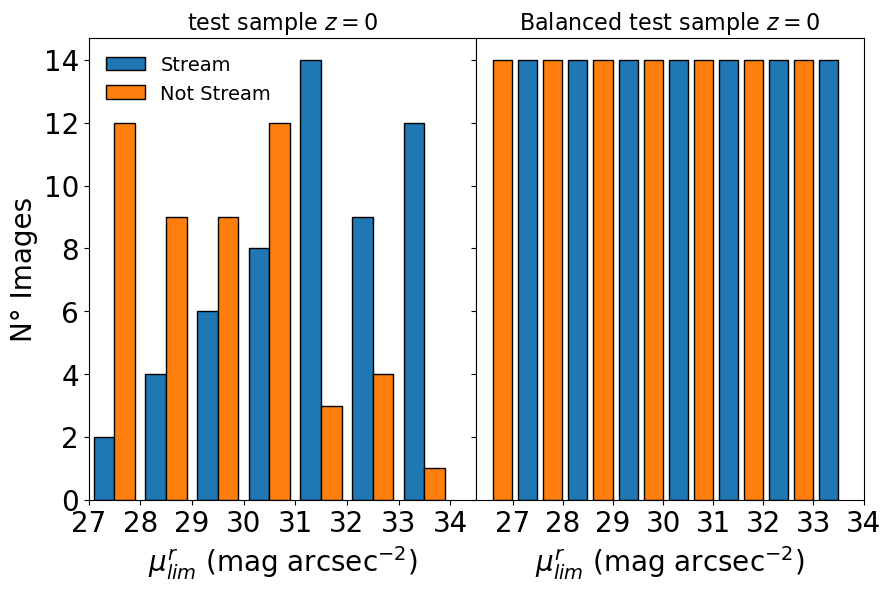}
    \caption{Example of data augmentation applied to the test sample of images with and without stellar streams. The left panel shows the original test sample distribution at $z=0$, where images are unevenly distributed across SB limits ($\mu_{lim}^{r}$) for stream and non-stream classifications. After the data augmentation, the right panel illustrates the balanced test sample, ensuring an equal number of images for each class across all surface brightness limits.}
    \label{fig:balanced}
\end{figure}

\begin{figure*}
    \centering
    \includegraphics[width=0.85\linewidth]{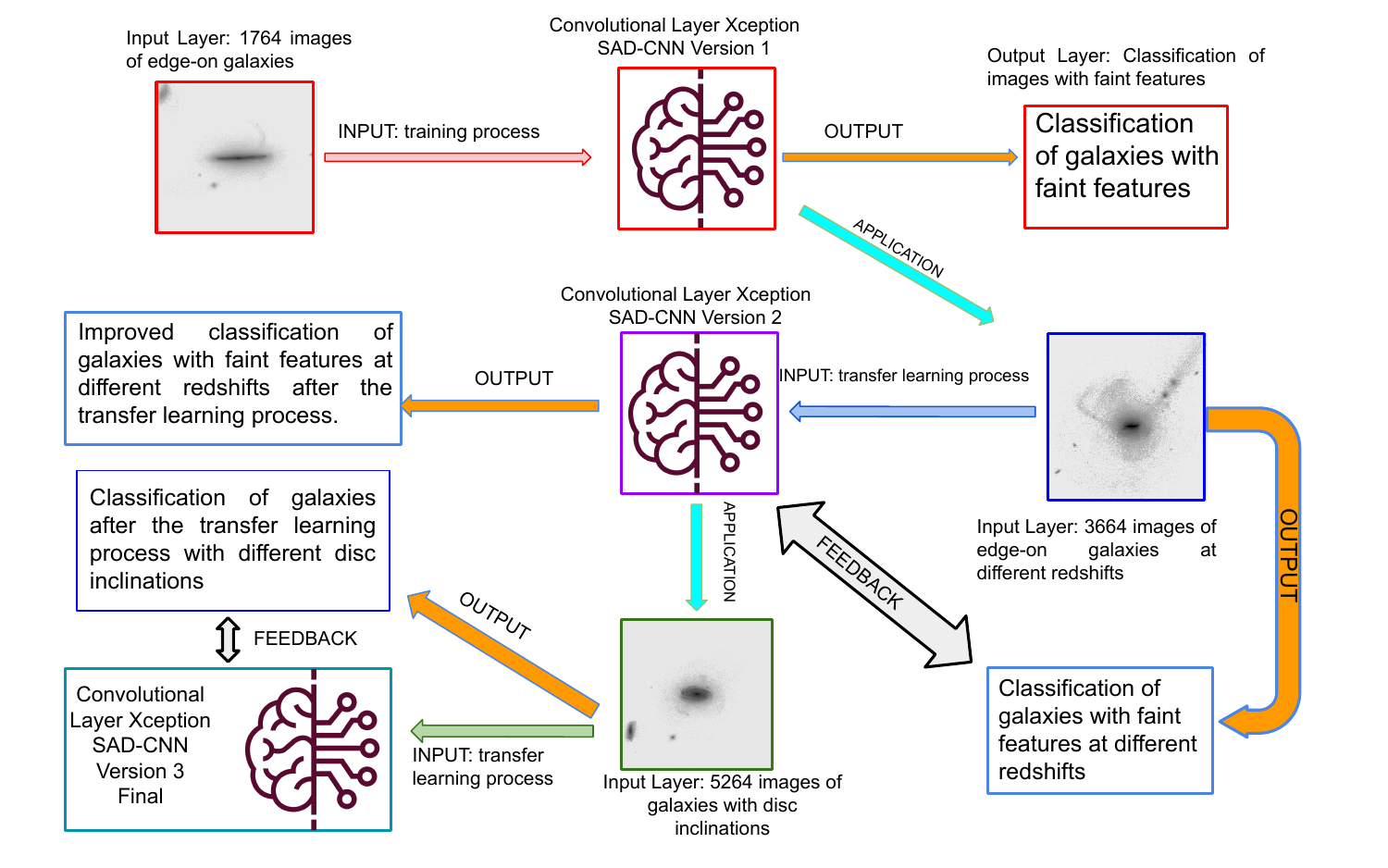}
    \caption{Overview of the iterative training and transfer learning process used in SAD-CNN. The first training stage begins with 1,764 images of edge-on galaxies at $z=0$, where the network learns to classify images based on the presence of faint features. The second stage introduces 3,664 additional images of edge-on galaxies at different redshifts, improving classification performance on galaxies observed at various cosmological distances. Finally, the third stage applies TL to a dataset of 5,264 images of galaxies with different disc inclinations, refining the model’s ability to generalize across varying orientations. Throughout the process, feedback is incorporated at each step, allowing for progressive learning and improved classification accuracy. }
    \label{fig:DiagCNN}
\end{figure*}

\section{THE AURIGA SIMULATIONS}
\label{sec:auriga}
The Auriga project is a suite of cosmological simulations designed to produce reasonably isolated galaxies in the mass range of the Milky Way \citep{AURIGA,2024arXiv240108750G}. These simulations consist of thirty (30) zoom halo simulations from the EAGLE project \citep{EAGLE}, performed in the framework of $\Lambda$CDM cosmology with parameters $\Omega_m$ = 0.307, $\Omega_b$ = 0.048, $\Omega_{\Lambda}$ = 0.693, and Hubble constant $H_0$ = 100 \textit{h} km s$^{-1}$ Mpc$^{-1}$, \textit{h} $= 0.6777$ \citep{param_cosmo_planck}. 
The simulations have a resolution of baryonic mass particles of $\sim 5 \times 10^4 M_{\odot}$ and resolution for dark matter particles $\sim 4 \times 10^5 M_{\odot}$, with a comoving softening length of 369 pc at $z=1$, after which the softening is kept constant in physical units. 
%The minimum comoving softening length of the gas cells follow that of the stellar particles. 
These correspond to "level 4" of Auriga Project. These simulations were run using the  {\sc arepo} code on a periodic cube with a side length of $100$ cMpc \citep{springel_2010,pakmor_2016}. {\sc arepo} solves the magnetohydrodynamical equations and integrates models of galaxy formation \citep{Vogelsberger2013, AURIGA}, incorporating sub-grid models for baryonic processes such as star formation \citep{Springel_Hernquist_2003}.

The Auriga model assigns a single stellar population to each stellar particle.  This assignment occurs every time there is a star formation episode, ensuring that each stellar particle is associated with a specific stellar population.
The properties of these stellar populations account for mass loss and chemical enrichment from Type Ia supernovae (SNIa) and asymptotic giant branch (AGB) stars, along with their respective ages and masses. These populations are characterized by their mass, age, mass-loss history, and chemical abundance patterns. 
The ISM is modelled using the subgrid multiphase approach introduced by \citet{Springel_Hernquist_2003}. It describes the dense and cold gas of the ISM with an effective equation of state that originates from the balance of gas cooling and heating by SN feedback.
%which regulates star formation and feedback through a self-regulated equilibrium model. Also stellar populations contribute mass and metals to the gas reservoir.
The photometric bands U, B, V, g, r, i, z, and K from \citet{BruzualCharlot} are calculated for each stellar particle, without accounting for dust extinction. 

The Auriga simulations are an excellent laboratory for studying the properties of stellar halos due to their high resolution \citep{AURIGA,2016MNRAS.459L..46M,2019MNRAS.485.2589M,2020MNRAS.497.4459F,2024arXiv241013491P}. In particular, \citet{2019MNRAS.485.2589M} presented a detailed comparison between simulated halos and those observed in the nearby Universe with the HST telescope within the Galaxy Halos, Outer disks, Substructure, Thick disks, and Star clusters (GHOSTS) project, showing that the parameters that characterize the Auriga stellar haloes, as well as their scatter, are generally in good agreement with the observed properties of nearby stellar haloes. \citet{2018MNRAS.478..548S}, showed that the luminosity function of satellites at redshift zero closely matches the observed luminosity function of both the Milky Way (MW) and Andromeda (M31). This agreement holds for satellites with stellar masses above 10$^{6} M_{\odot}$ , which aligns with the resolution used in these studies. 
%\citet{Vera-casanova2022} demonstrated that a significant fraction of stellar streams generated by satellites is present in the majority of models at redshift zero, with a wide range of SB limits. However, these streams do not necessarily originate from the most massive accretion events. 
\citet{Vera-casanova2022} showed that approximately $87\%$ of the simulated halos exhibit stellar streams at redshift $z=0$, across a wide range of surface brightness limits. However, these streams do not necessarily originate from the most massive accretion events, as even satellites disrupted several gigayears ago can leave detectable tidal features that persist until the present time.
\citet{2024arXiv241009144R} and \citet{Shipp2024} further explored the disruption of satellite galaxies around the Auriga haloes. They 
found that the distribution of streams in pericentre-apocentre space 
 overlaps significantly with the Milky Way intact satellite population. However their results suggest that either cosmological simulations, such as Auriga, are disrupting satellites far too readily, or that the Milky Way’s satellites are more disrupted than current imaging surveys have revealed.

\section{METHODOLOGY}
\label{sec:meth}
In this work, we use the Auriga simulations to generate surface brightness maps of late-type galaxies at different redshifts, inclinations, and surface brightness limits. These images are used to train a convolutional neural network (CNN) to rapidly and efficiently identify low surface brightness features. Our training is done in stages, increasing the complexity behind the detectability of these features at every step. As a result, we seek to understand the main \rm limitations behind this type of approach. Our training sets consist of $\sim 10,000$ different images with different projections and rotations of galaxies. In this sample, we indicated the presence or absence of a stellar stream by visual inspection of the images (see Sec. \ref{SBM}).

\subsection{Surface Brightness Maps}
\label{SBM}
Each stellar particle in the Auriga simulation represents a different stellar population, with a given mass, age, and metallicity. As discussed in Sec. \ref{sec:auriga}, each particle has determined its luminosity in the U, V, B, K, g, r, i, and z bands, calculated based on its mass, age, and metallicity. These luminosities are derived following the methodology described in \citet{BruzualCharlot}.
To mitigate the effects of dust extinction, and to have a better tracer of the underlying mass distribution, previous observational works \citep[e.g.,][]{2013ApJ...765...28A,2018A&A...614A.143M,MartinLSST2022,Juan_MiroCarretero2023,Martínez-Delgado2023,2023ApJ...949L..37S}, have focused their analysis on SB maps obtained in the photometric $r-$band. Auriga does not include a self-consistent dust model, meaning that these luminosities are computed without the direct effects of dust attenuation. Thus,  for the CNN training, we employed SB Maps derived from the luminosity distribution of each simulated galaxy in this band. We considered all stellar particles located within a radial distance of 150 kpc from the galactic centre. In this work, the galactic center is defined following \citet{Vera-casanova2022}, where it is determined by identifying the particle most bound to the system.

Building on the detection rates reported in \citet{Vera-casanova2022} and \citet{2024arXiv240903585M}, we note that stellar streams become increasingly prominent for surface brightness limits $\geq 27$ (mag arcsec$^{-2}$). Based on this, we generate SB maps of our Auriga models considering a limiting magnitude range from 27 to 34 mag arcsec$^{-2}$, and steps of 1 mag arcsec$^{-2}$.
By systematically varying the limiting magnitudes, we approach different observing depths and evaluate the performance of our CNN over a range of surface brightness limits.
To assure uniformity and comparability between the generated SB maps, we performed a normalisation process before using them to train the CNN. Each map was converted into a FITS image format and then normalised by its maximum pixel luminosity value, resulting in pixel values ranging from zero to one. The normalisation procedure effectively standardises the intensity distribution across all maps, independent of the specific limiting magnitude. This strategy allows us to mitigate biases due to differences in magnitudes in the analysed SB maps and ensures that the CNN learns to discern faint substructures rather than being influenced by the typical low SB values associated with these substructures. In addition, all images were generated considering 300$\times$300 pixels size. This corresponds to a physical resolution of 1 kpc per pixel. This standardised resolution facilitates consistency in feature recognition and analysis across the entire dataset. We note that, when mocking observations from different instruments, the spatial resolution and associated observational errors may affect the detectability of some of the faintest features that exist \citep{2024arXiv240903585M}. This, nevertheless, does not affect the results of our training process. 

The generated SB maps were utilized to identify low surface brightness features (LSBF) in each galaxy. This was done by visual inspection of each image, following the procedure described in \citet{Vera-casanova2022}. Specifically, images without any discernible features were assigned a value of zero, while images exhibiting faint structures were assigned a value of one. During this procedure we decided to not classify low SB satellites that still gravitationally bound as faint features. For this procedure, we made individual selections between co-authors, followed by multiple iterations and discussions over the selection. This collaborative process eventually culminated in a shared visual inspection.

\begin{figure*}
    \centering
    \includegraphics[width=1\linewidth]{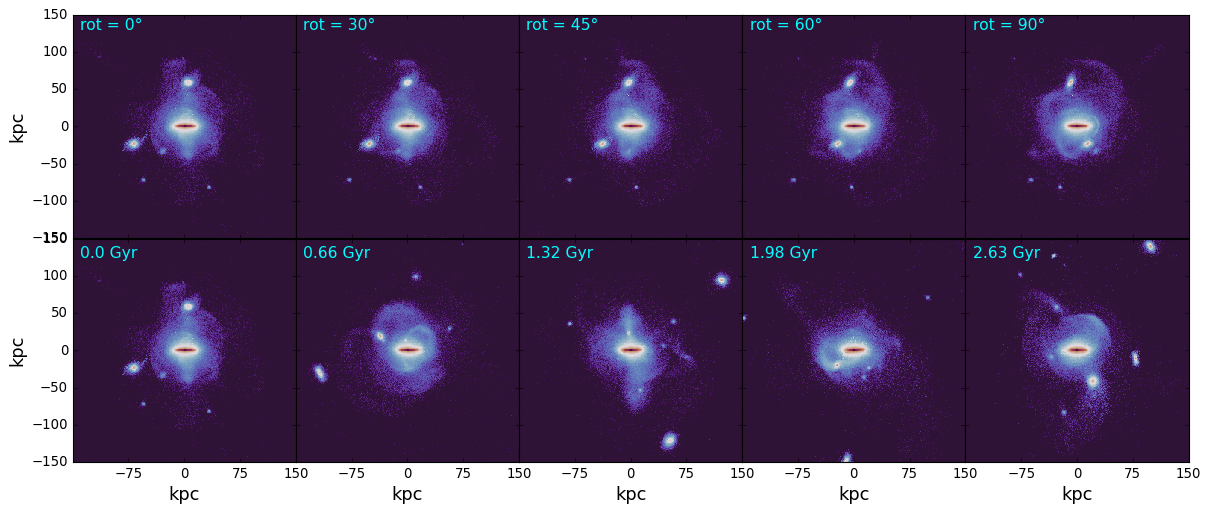}
    \caption{Surface brightness maps of halo 12 from the Auriga simulation (AU12), illustrating the evolution of stellar streams under different disc plane rotations and snapshot times. The first row shows how the shape of the stream changes due to different viewing angles, while the second row highlights the temporal evolution of the streams. The maps have a surface brightness limit of $\mu_{lim}^{r} = 32$ mag arcsec$^{-2}$ and a box of 300 kpc by side.}
    \label{fig:example_halos}
\end{figure*}

\begin{figure}
    \centering
    \includegraphics[width=0.49\textwidth]{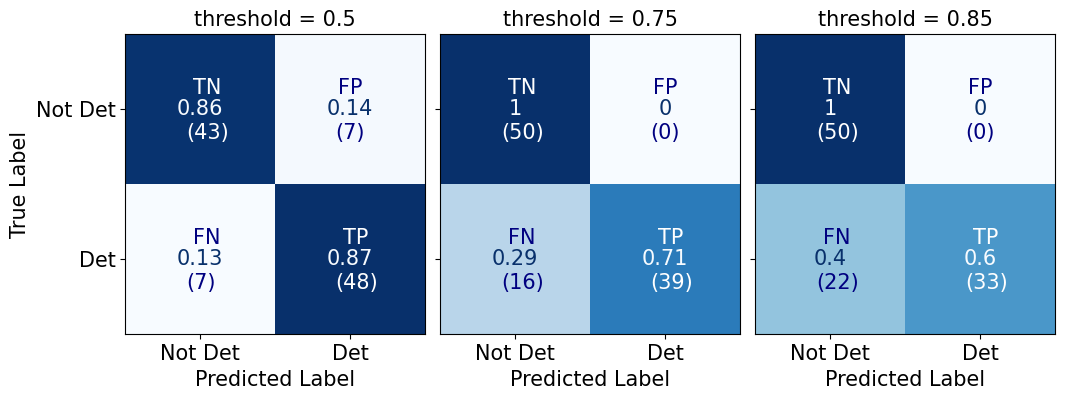}
    \caption{The panels display confusion matrices for the test set of galaxies at redshift zero, evaluated with threshold values of 0.5 (left), 0.75 (middle), and 0.85 (right). Each matrix corresponds to a subset of 105 test images. The top-left cell represents the true negative rate, the top-right cell the false positive rate, the bottom-left cell the false negative rate, and the bottom-right cell the true positive rate alongside the absolute counts (in parentheses). Further details on the performance metrics are provided in Table \ref{tab:tab_statistic1}.}
    \label{fig:Matrixz0}
\end{figure}

\subsection{Deep Learning}
\label{CNN}

A CNN employs a convolutional operator to process the images, by applying a discrete linear transformation over the input data matrix. In this case, the input data consists of images representing SB maps of our simulated galaxies. CNNs are typically composed of four fundamental components: convolutional layers, regularisation layers, clustering layers, and fully connected layers. These layers work together to extract meaningful features from the input data, and the final layer uses these features to generate predictions. Regularization layers, such as the exclusion layer, are used to improve network generalization and mitigate overfitting by randomly excluding neurons during training. Through a training process, the network learns to capture non-linear combinations of features, allowing accurate predictions to be made on new input data. As previously stated, the CNN applies a discrete linear transformation over the input image and provides weights for every discrete position of the corresponding matrix. During this process it applies multiple combinations of different kernels to capture different features and patterns in the input images, helping the classification process \citep[see][]{dumoulin2016guide}. The input data for the convolutional layers consist of a three dimensional matrix, two spatial dimensions, and a third dimension with intensity values. This last dimension is usually called the dimension channel, which has information on quantities such as colors or brightness intensities.

Here we utilise the Xception \citep{2016arXiv161002357C} CNN architecture to identify low SB features on mock galaxy images. The Xception model utilises Depthwise Separable Convolution also called "separable convolution", a technique that factorizes standard convolutions into two separate operations: a depthwise convolution, which applies spatial filtering independently to each input channel, followed by a pointwise convolution, which combines information across channels. This approach reduces computational cost while maintaining high classification performance. This approach is computationally efficient and helps reduce the number of parameters in the network. The design choice allows Xception to optimise model size, computational complexity, and performance.  In this work, we use a sigmoid function output to deliver one value per class for each of the examples. The output layer scales the network output between 0 and 1. This allows us to interpret the results as the probability that a given image contains a feature learned during the training process. For a more detailed explanation, see \citet{2016arXiv161002357C}.

\begin{figure}
    \centering
     \includegraphics[width=0.49\textwidth]{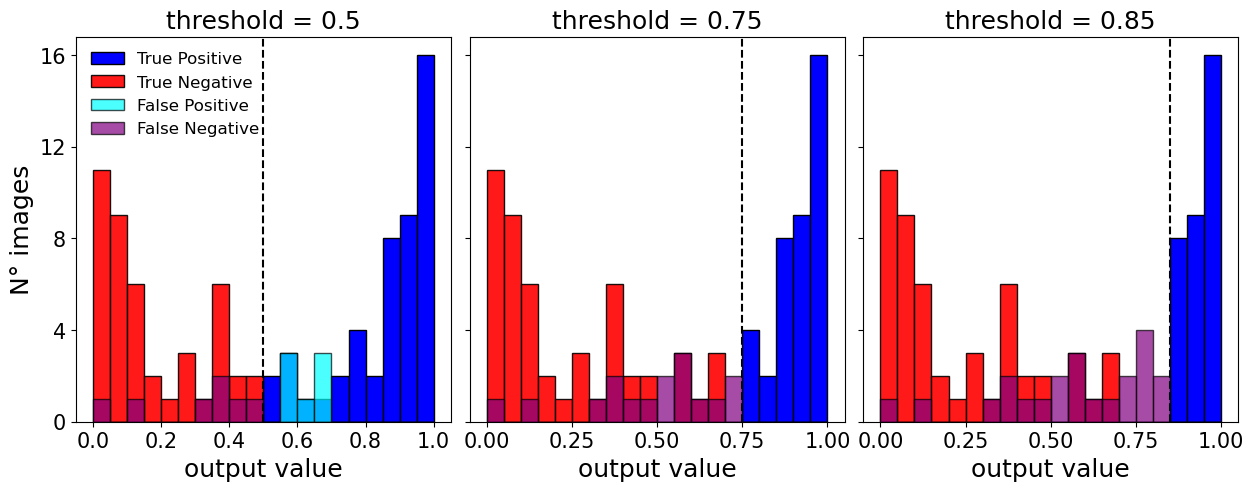}
    \caption{Distribution of the prediction of SAD-CNN in the galaxies with edge-on projection at \textit{z=0} for different threshold values introduced by Figure. \ref{fig:Matrixz0}. The threshold level is indicated with the dashed line. The red distribution shows the true negative cases. The purple distribution shows the false negative cases. The Blue distribution shows the true positive cases. The cyan distribution shows the False positive cases by the SAD-CNN.}
    \label{fig:Hist_z0}
\end{figure}

\subsection{Training method and metrics}
\label{training}

The training process consists of a minimisation problem to fit the data. As previously discussed, during this process different kernels, and their associated weights, are considered to find the optimal configuration that best summarises and classifies the images. A common approach is to initialise the weights with random values and then, through a trial and error process, optimise these weights to recognise the images. To conduct the trial and error process, we need to divide the data into three independent sets: Training set, Validation set, and Test set. We allocated 80 percent, 10 percent, and 10 percent of the total sample of images to each set, respectively. The training set is used directly to adjust the network's weights, improving classification through an iterative process. The validation set was used during training to indirectly evaluate the performance of the model and adjust the parameters as needed, while the test set, composed of previously unseen images, is reserved for the final evaluation of the accuracy of the model.

To guarantee the robustness of our model, we have balanced the data set so that it has an equal number of images with and without LSBF at each surface brightness limit. This balance is key to prevent biases in the training process and to ensure that the model learns to accurately distinguish between the two classes under varying conditions. To achieve this, we applied a data augmentation process which consist of performing simple transformations on the images, such as rotations and reflections. In this way, we increased the diversity of the training data, which helped the model to better generalize the unseen data. Figure \ref{fig:balanced} shows an example of this procedure.  Importantly, the mock images selected to perform the data augmentation process are randomly selected. This approach helps mitigate the natural bias introduced by galaxies observed at deeper SB limits, ensuring a more accurate evaluation of the ability of the model to detect LSBF in different scenarios.

The model classifies an image as containing an LSBF if the predicted probability exceeds a threshold of 0.5; otherwise, it is classified as not containing such features. To assess the classification performance, we define four possible outcomes based on visual classification: True Positives (TP): The network correctly identifies the presence of an LSBF. False Positives (FP): The network incorrectly classifies an image as containing an LSBF when it does not. True Negatives (TN): The network correctly identifies the absence of an LSBF. False Negatives (FN): The network fails to detect an LSBF that is actually present.
The results of the trained network are evaluated using various performance metrics, which are quantitative measures used to assess performance on specific tasks. They are also used by the model to help guide the iterative process of improving model design and performance. Common metrics include accuracy, precision, recall, and the F1-score. Our primary metric for evaluation, accuracy, is calculated as:
\begin{equation}
    \textbf{accuracy} = \frac{\textbf{true positives + true negatives}}{\textbf{(All cases)}}.
    \label{eq:acc}
\end{equation}
Accuracy measures the percentage of instances correctly classified from the total dataset and it is commonly used in classification tasks. Precision, defined as
\begin{equation}
    \textbf{Precision} = \frac{\textbf{true positives}}{\textbf{(true positives + false positives)}},
    \label{eq:prec}
\end{equation}
represents the ratio of true positive predictions to all
positive predictions.  Recall
\begin{equation}
    \textbf{Recall} = \frac{\textbf{true positives}}{\textbf{(true positives + false negatives)}}
    \label{eq:recall}
\end{equation}
 measures the proportion of true positive predictions out of all actual positive instances in the dataset.

Finally, we also use the F1-score metric, which combines precision and recall into a single value, 
\begin{equation}
    \textbf{F1-score} = \frac{1}{2} \times \frac{(\textbf{precision} * \textbf{recall})}{(\textbf{precision + recall})}.
    \label{eq:f1}
\end{equation}
The F1-score provides a balanced assessment of the performance model, especially in situations where precision and recall need to be balanced. It is defined as the harmonic mean of precision and recall, ranging from $0$ to $1$, where $1$ indicates perfect precision and recall, and $0$ indicates that the model has completely failed to identify positive cases.

In this work, we apply a technique known as Transfer Learning (TL), which involves reusing knowledge learned from one dataset to improve performance on a related but different dataset. For instance, in the context of galaxy morphological classification, weights trained on astronomical images can serve as a starting point. Additionally, studies have demonstrated that TL can even be effective when using weights pre-trained on unrelated datasets \citep[e.g.,][]{2022arXiv220109679I}. In our case, we initialize our model with weights pre-trained on the ImageNet dataset \citep{5206848}, which contains a broad range of labelled images across various categories. This approach allows the network to start with a general understanding of image features, improving the efficiency of training on our specific task.
To train the network, we use the Adam optimizer \citep{2014arXiv1412.6980K} with a learning rate of 0.0001, running the model for up to 200 epochs. To prevent overfitting, we employ early stopping. This method monitors a selected metric (such as validation loss) and halts training if no improvement is observed within a specified number of epochs\footnote{Learning curves are shown in the \hyperref[sec:app]{Appendix.}} \citep{team2022keras}.
When incorporating new data into our analysis, such as images of galaxies with varying inclinations, we concatenate a random 20 percent of the previous training set with the next one and retain the pre-trained weights from the earlier training stages. This approach ensures that the model builds on previously learned features while adapting to the additional data. The resulting network combines the weights from previous training sessions with the new dataset, providing a robust model capable of generalizing to more diverse conditions.

\begin{figure*}
    \centering
    \includegraphics[width=1\textwidth]{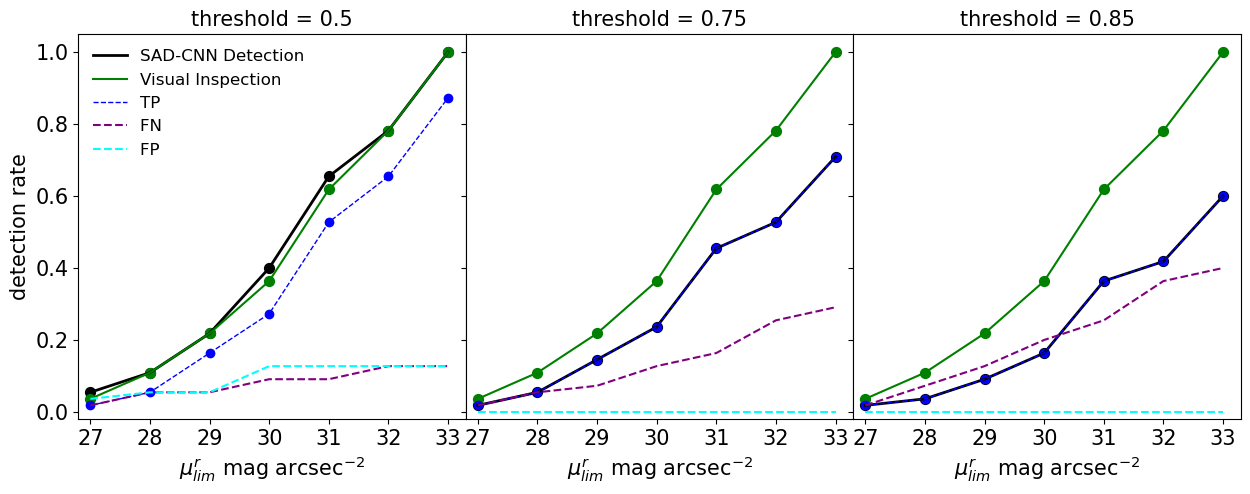}
    \caption{Cumulative detection rate of LSBF for all galaxies at $z=0$. The black line represents the cumulative detection rate by SAD-CNN, while the green line corresponds to cumulative detection rate from visual inspection. The blue dashed line shows the cumulative fraction of only the true positive cases. The purple dashed line is the cumulative fraction rate produced with false negative cases, while the cyan dashed line corresponds to false positive detections. An over-detection trend is noticeable at SB limits fainter than $\approx$ 29 mag arcsec$^{-2}$ which remains consistent across different threshold levels but is slightly shifted depending on the chosen threshold.}

    \label{fig:cumulative_z0}
\end{figure*}

\subsection{Training sets definition}
\label{groups}
Our training methodology follows a stepwise approach, where the model is first trained on a simpler dataset and then progressively exposed to more complex cases.
%involves a process of progressive refinement, gradually expanding to encompass a wider range of images with different configurations.
At each training step, the effectiveness of the trained model was evaluated using a test set. As a first step, we trained the model only considering edge-on galaxies at $z=0$. This is the simplest possible configuration as it avoids issues associated with the extended light distribution of the discs and their internal structure. Once tested, the CNN was further trained and tested using a large sample of edge-on galaxies obtained at different redshifts. We finally generated a sample of late-type galaxies with different inclinations. This allowed us to test the performance of our previously trained network on this more challenging configuration. Based on this evaluation, we refined the model by incorporating images of inclined galaxies into the training process, which allowed it to generalise better to different orientations. A summary of our training procedure is presented in the scheme of Figure \ref{fig:DiagCNN}.

To carry out this iterative procedure, we generated different data sets. These sets are described below.

\begin{itemize}
    \item Group I: Comprises 210 images of our simulated galaxies in an edge-on projection, all at $z=0$. We remind the reader that, from each Auriga model, several images are obtained by systematically varying the limiting SB magnitude. These images serve as the foundation for the initial training phase. We consider this to be the most simple case for the network, as we can observe the halo without having the disc covering a large area of the image. It also eliminates potential problems with disc features, such as spiral arms. This group encompassed variations of edge-on galaxies achieved through rotations about the symmetry axis of our disc galaxies. Specifically, rotations of $30^{\circ}$, $45^{\circ}$, $60^{\circ}$, and $90^{\circ}$ were applied to the original 210 images, resulting in a total of 1050 distinct FITS images. 
    Each rotation introduced unique configurations of the stellar streams and satellites in configuration space, as shown in Figure \ref{fig:example_halos}. In particular, some rotations, such as the $45^{\circ}$ rotation depicted in the middle column of the lower row, led to scenarios where the stellar streams are hard to disentangle from the stellar halo component. After the data augmentation, the number of images was raised to 1,764. To create the test set, we randomly selected 10 percent of these images. %Contrary to the data augmentation process, this rotation provide entirely new spatial configurations of debris and thus, which can thus be regarded as independent data sets. 

    \item Group II: In the subsequent learning phase, a new sample of images was introduced, comprising galaxies within the following redshifts; $z=$ [0.024, 0.049, 0.074, 0.099, 0.126, 0.153, 0.180, 0.214, 0.244, 0.276].  This group consisted of 2100 images, representing galaxies at different stages of late evolution, influenced by their merger history. An example of such images is shown in Figure \ref{fig:example_halos} using the same halo rotated previously. Ten percent of this sample was separated into a test set. After the separation, we augmented the data by applying rotations and random flips to the images, resulting in 3,664 images.
    
    \item Group III: Finally, the last learning phase introduced galaxies with varying inclinations of the disc symmetry axis for the line of sight. This group incorporated inclinations of 15$^{\circ}$, 30$^{\circ}$, and 45$^{\circ}$ of the images from Group I, resulting in an additional 3150 images. Ten percent of each inclination was separated into a test set. After the data augmentation process, we obtain 5,264 images. This new set allowed CNN to refine its ability to distinguish between different structural features, particularly in the presence of spiral arms and tidal features.
\end{itemize}

\section{Training Process}

We now discuss the process carried out to systematically and gradually train our CNN using the different samples and metrics discussed in Sec. \ref{sec:meth}. Our goal is to provide a more diverse and complex data set at each training step, thus learning about the limitations behind this kind of method, and also providing potential ideas for future efforts.

\begin{figure}
    \centering
    \includegraphics[width=1\linewidth]{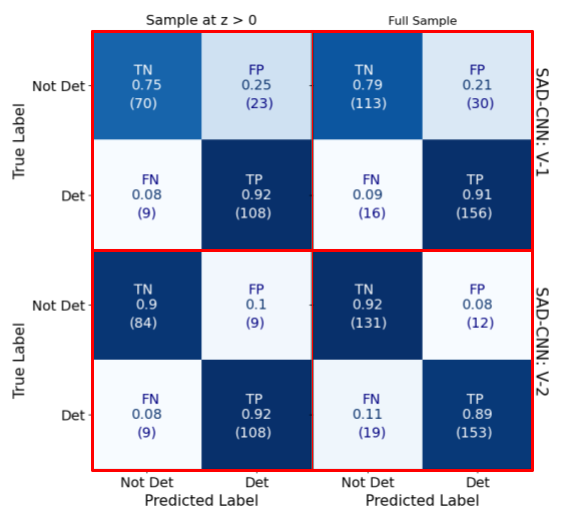}
    \caption{Confusion matrices for SAD-CNN versions V1 and V2. The matrices on the left represent performance on the test sample at $z>0$, while those on the right combine results from test samples. True negative (TN), false positive (FP), false negative (FN), and true positive (TP) rates are shown alongside the absolute counts (in parentheses). SAD-CNN V2 demonstrates improved performance, particularly in reducing false positives (FP), as seen in the balanced predictions for both test scenarios. }
    \label{fig:matrix_zdif}
\end{figure}

\subsection{Edge-on Galaxies at Redshift zero}

For our initial training, we used the dataset from Group I. As mentioned in Section \ref{training}, this initial dataset covered 1,764 (1050 originals) models of galaxies in an edge-on projection at $z=0$. For the training, validation, and test sets we randomly assign 80, 10, and 10 percent of the images, respectively. As previously discussed, the output of the CNN for each image is a value ranging from 0 to 1, which can be treated as the probability of the given images containing a LSBF. Values close to one indicate a high probability that a faint feature is present in the image, while values close to zero suggest a low probability.

The statistical summary of our first training step is presented in Table \ref{tab:tab_statistic1}. In the initial phase, we consider outputs of the CNN above a threshold value of 0.5 as LSBF detections. With this threshold, our CNN achieved a precision of 0.87 in correctly identifying images without LSBF and 0.87 in detecting images with LSBF. Averaging, we obtained an overall precision, recall and F1-score of  0.87, indicating a well balanced performance assessment of our model. Following the initial training, we explored different threshold values to adjust the compromise between completeness and accuracy in stream detection. For example, a higher threshold value would produce a purer sample of true positives. However, this would result in a larger misidentification of images with streams as negative detections. To test this, we considered thresholds of 0.75 and 0.85. The results are also shown in Table \ref{tab:tab_statistic1}. As expected, with these higher thresholds we recovered a purer true positive sample, but we misidentified a larger fraction of images with low surface brightness features as negative detection. This can be seen in the lower values of the Recall and F1-scores. 

Figure \ref{fig:Matrixz0} presents the confusion matrix obtained in this phase. The confusion matrix is a tabular representation that summarises the performance of a classification model, detailing metrics such as true positive rate (TP, equivalent to recall of detections), false positive rate (FP), true negative rate (TN, equivalent to recall of not detections), and false negative rate (FN). The left panel shows the matrix obtained with a 0.5 threshold value, where we obtain a reasonably high rate of TP and TN cases, both above 86 percent. The middle and right panels show the matrices obtained for threshold values of 0.75 and 0.85, respectively. As the threshold increased, the rates of FP and TP cases fell, while the rates of TN and FN cases increased. Note how the TP detection falls from 0.87 to 0.6 with the higher considered threshold. This exercise clearly shows that, while higher thresholds allow us to obtain a much cleaner sample of images with stellar streams (very low values of FP), this choice also results in a significant misclassification of images with streams as non-detections.

\begin{table}
    \centering
    %\scriptsize
    \resizebox{\columnwidth}{!}{
    \begin{tabular}{|l|l|l l l|l|}
    \hline
    \multicolumn{6}{ |c| }{SAD-CNN: Version-1} \\
    \hline
    Threshold & Label & Precision & Recall & F1-score & support\\
        \hline
          & Not Det  & 0.86 & 0.86 & 0.86 & 50 \\
    0.5   & Det     & 0.87 & 0.87 & 0.87 & 55 \\
          & average & 0.87 & 0.87 & 0.87 & 105 \\
        \cline{2-6}
          & accuracy & & & 0.87 & 105\\
        \hline
          & Not Det  & 0.76 & 1.00 & 0.86 & 50 \\
    0.75  & Det     & 1.00 & 0.71 & 0.83 & 55 \\
          & average & 0.88 & 0.85 & 0.85 & 105 \\
        \cline{2-6}
          & accuracy & & & 0.85 & 105\\
        \hline
          & Not Det  & 0.69 & 1.00 & 0.82 & 50 \\
    0.85  & Det     & 1.00 & 0.60 & 0.75 & 55 \\
          & average & 0.85 & 0.80  & 0.78 & 105 \\
        \cline{2-6} 
          & accuracy & & & 0.79  & 105\\
        \hline
    \end{tabular}
    }
    
    \caption{The summary of the values at different threshold values obtained. The metrics are defined in the section \ref{CNN}. The accuracy of the different threshold levels is put at the bottom of each case. The support represents the quantity of images used to evaluate this metric. The statistic corresponds to the test set of data Group I, with 105 different images edge-on projection at $z=0$. }    
    \label{tab:tab_statistic1}
\end{table}

We now explore in detail the distribution of classification model outputs, and how they correlate with our visual classification. This is shown in Figure \ref{fig:Hist_z0}, where red and blue histograms show the output value distribution of TN and TP cases, respectively. Different panels correspond to the different thresholds considered. The histograms in this figure show a biased distribution of true cases towards 0 and 1, indicating that for a significant number of cases the CNN output aligns well with our visual classification. The FP and FN cases are shown in cyan and purple, respectively. Notably, false detections occur throughout the CNN output range, but are more frequent near the 0.5 threshold. Increasing the threshold significantly reduces the number of false positive detections (cyan histograms), in this particular case to 0, but increases false negative cases (purple histograms). Consequently, for very high precision values, several galaxies with visually identified streams are misclassified as non-detection cases.

We finally summarise the classification processes in Figure \ref{fig:cumulative_z0}, where we show the normalised cumulative function of images with detected LSBF as a function of limiting SB magnitude, considering the threshold values previously introduced. The left, middle, and right panels correspond to precision thresholds of 0.5, 0,75 and 0.85, respectively. In each panel, the black and green lines show the identification obtained by SAD-CNN and our visual inspection, respectively. The blue line represents the cumulative functions obtained considering only TP (detections in agreement with our visual inspection).
For the canonical threshold of 0.5  (left panel), the detection rate obtained by the CNN (black line), as a function of  $\mu_{lim}$, closely matches the cumulative function obtained from the visual inspection (green line). This could be due to a mix of true and false positive detections adding up to a similar cumulative function. However, as shown by the blue line, the cumulative function of  true positive cases closely follows the CNN detection curve,  highlighting the excellent performance of the model. This can also be seen by a very low fraction of false positive and negative detections at all $\mu_{lim}$, shown by the purple and cyan lines. In the middle panel, by setting the threshold value at 0.75, we observe a decrease in the CNN detection rate (black line). However, it is now perfectly followed by the cumulative function of true positive cases, indicating a purer classification. Unfortunately, as previously discussed, this larger threshold is also associated with a significant increase in False negative cases (see purple line). Finally, in the right panel, we show the results obtained with a threshold equal to 0.85. Note that the purity of the CNN detection does not improve and, at the same time, the number of False negative cases significantly increases. As a result of this analysis, from now on, we will work with a threshold of 0.5.

\subsection{Edge-on Galaxies at different times}

The morphology of galaxies, and their stellar halos, can evolve rapidly due to processes such as accretion and mergers. As a result, it is possible to use the same simulated galaxies at different redshifts as a training set. Also the orientation often changes quickly as well. The images generated for Group II comprise halos at different snapshot times, with intervals of $\sim$ 0.25 Gyr, up to a lookback time of 3 Gyr. Figure \ref{fig:example_halos} (bottom panels)  illustrates one example of halo with recent significant accretions. As the corresponding satellites orbit their host and suffer tidal disruption, they deposit material in different regions of the host halo, thus altering the structure of the resulting image. Additionally, the earlier tidal streams have time to evolve. Within this set, we added 3664 (2100 originals) distinct images that the CNN has not previously analysed during training. As already mentioned, the authors label all samples to indicate the presence or absence of LSBF, and, subsequently, we perform a data augmentation process and apply TL.

\begin{figure}
    \centering
    \includegraphics[width=1\linewidth]{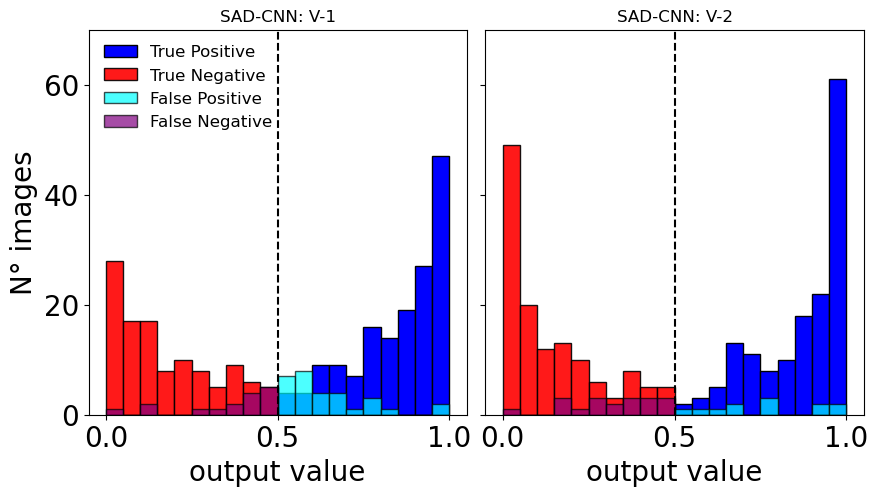}
    \caption{Distribution of the prediction values of the first and second versions of the SAD-CNN in each panel respectively. The threshold value used in these distributions is 0.5 indicated by the dashed line. In both cases the values in both test sets of Groups I and II. The red distribution shows all true negative cases for each SAD-CNN version. The purple distribution shows the false negative cases. The blue distribution shows all detection by the SAD-CNN or the true positive. False positives are present by the cyan distribution.}
    \label{fig:Hist_zdif}
\end{figure}

The results of this new training process are shown in Figure \ref{fig:matrix_zdif}, where we compare the CNN performance in the Group II test set, before and after applying the TL procedure. The two top and two bottom panels show the confusion matrix obtained for the previously trained version of SAD-CNN (V-1 hereafter), and the version obtained after TL (V-2 hereafter), respectively.
When considering only images obtained at $z>0$ (left panels) we observe that while V-1 was producing a poor classification of the Group II test images, after the TL procedure (bottom middle) the results significantly improved. In particular, we notice that the fraction of false positive decreased, improving the purity of the true positive sample. Note that the F1-score increases from an average 0.84 in the V-1 to 0.91 in the V-2 version. The right panels show the results obtained using the combined test sample from Group I and II; i.e. with images at all z considered. Both versions are consistent. Again, we notice a significant improvement of the metrics, with an average F1-score increasing from an average 0.85 in the V-1 to 0.90 in the V-2 version.

The left and right panels of Figure \ref{fig:Hist_zdif} show the distribution of output values obtained from the V-1 and V-2 CNN versions, respectively. Here we focus on the "all $z$" sample. We note that even though the distributions are similar, the V-2 has a more skewed distribution towards the maximum and minimum output values. This indicates that thanks to further training, CNN is more certain about its assessment. Table \ref{tab:tab_statistic3} summarizes the results of this second training process. 

\begin{table}
    \centering
    \resizebox{\columnwidth}{!}{
    \begin{tabular}{|l|l|l|l|l|l|}
    \hline
    \multicolumn{6}{ |c| }{SAD-CNN: Version-1} \\
    \hline
        Redshift & Label & Precision & Recall & F1-score & Supp\\
         \hline
        & Not Det  & 0.89 & 0.75 & 0.81 & 93 \\
      $z>0$ & Det & 0.82 & 0.92 & 0.87 & 117 \\
        & average & 0.86 & 0.84 & 0.84 & 210 \\
        \cline{2-6}
        & accuracy & & & 0.85 & 210\\
        \hline
        & Not Det  & 0.88 & 0.79 & 0.83 & 143 \\
      all $z$ & Det & 0.84 & 0.91 & 0.87 & 172 \\
        & average & 0.86 & 0.85 & 0.85 & 315 \\
        \cline{2-6}
        & accuracy & & & 0.85 & 315 \\
        \hline
    \multicolumn{6}{ |c| }{SAD-CNN: Version-2} \\
    \hline
        Redshift & Label & Precision & Recall & F1-score & Supp\\
         \hline
        & Not Det  & 0.90 & 0.90 & 0.90 & 93 \\
      $z>0$ & Det & 0.92 & 0.92 & 0.92 & 117 \\
        & average & 0.91 & 0.91 & 0.91 & 210 \\
        \cline{2-6}
        & accuracy & & & 0.91 & 210\\
        \hline
        & Not Det  & 0.87 & 0.92 & 0.89 & 143 \\
      all $z$ & Det & 0.93 & 0.89 & 0.91 & 172 \\
        & average & 0.90 & 0.90 & 0.90 & 315 \\
        \cline{2-6}
        & accuracy & & & 0.90 & 315 \\
        \hline

    \end{tabular}
    }
    \caption{Summary of the metrics values of the two version of our CNN model for different groups of galaxies. Group I and Group II finally combined both test samples in all redshift. All statistics are obtained with a threshold of 0.5 value. The metrics are defined in the section \ref{CNN}. The two sample values correspond to the test set of the corresponding dataset.}    
    \label{tab:tab_statistic3}
\end{table}

\begin{table*}
    \centering
    \begin{tabular}{|c|c|c|c|c|c|c|c|c|}
        \hline
        &  & \multicolumn{3}{|c|}{SAD-CNN: Version-2} & \multicolumn{3}{|c|}{SAD-CNN: Version-3} & \\
        \cline{3-8}
        Inclination & Metric & Precision & Recall & F1-score & Precision & Recall & F1-score & support \\
        \hline
        \multirow{4}{*}{15$^{\circ}$}
        & Not Det   & 0.84 & 0.76 & 0.80 & 0.81 & 0.81 & 0.81 & 42 \\
        & Det      & 0.86 & 0.91 & 0.89 & 0.88 & 0.88 & 0.88 & 68 \\
        & Average  & 0.85 & 0.84 & 0.84 & 0.85 & 0.85 & 0.85 & 110 \\
        & Accuracy &      &      & 0.85 &      &      & 0.85 & 110 \\
        \hline
        \multirow{4}{*}{30$^{\circ}$}
        & Not Det   & 0.73 & 0.58 & 0.65 & 0.72 & 0.89 & 0.80 & 38 \\
        & Det      & 0.77 & 0.87 & 0.82 & 0.92 & 0.79 & 0.85 & 62 \\
        & Average  & 0.75 & 0.72 & 0.73 & 0.82 & 0.84 & 0.83 & 100 \\
        & Accuracy &      &      & 0.76 &      &      & 0.83 & 100 \\
        \hline
        \multirow{4}{*}{45$^{\circ}$}
        & Not Det   & 1.00 & 0.30 & 0.46 & 0.84 & 0.74 & 0.78 & 42 \\
        & Det      & 0.66 & 1.00 & 0.80 & 0.84 & 0.90 & 0.87 & 63 \\
        & Average  & 0.83 & 0.65 & 0.63 & 0.84 & 0.82 & 0.83 & 105 \\
        & Accuracy &      &      & 0.70 &      &      & 0.84 & 105 \\
        \hline
        \multirow{4}{*}{All}
        & Not Det   & 0.83 & 0.54 & 0.65 & 0.79 & 0.82 & 0.80 & 122 \\
        & Det      & 0.76 & 0.93 & 0.83 & 0.88 & 0.86 & 0.87 & 193 \\
        & Average  & 0.79 & 0.73 & 0.74 & 0.84 & 0.84 & 0.84 & 315 \\
        & Accuracy &      &      & 0.77 &      &      & 0.84 & 315 \\
        \hline
    \end{tabular}
    
    \caption{Comparison of SAD-CNN metrics for Version-2 and Version-3 at different inclinations (15$^{\circ}$, 30$^{\circ}$, 45$^{\circ}$) and overall (All). Metrics include Precision, Recall, F1-score, and Accuracy for each configuration. The metrics obtained for the test set were 315 images of group III with a total of 4500 images at different inclinations. The values were obtained with a threshold value of 0.5. }
    \label{tab:tab_statistic4}
\end{table*}

\subsubsection{Most Common False Cases}

\begin{figure}
    \centering
    \includegraphics[width=0.5\textwidth]{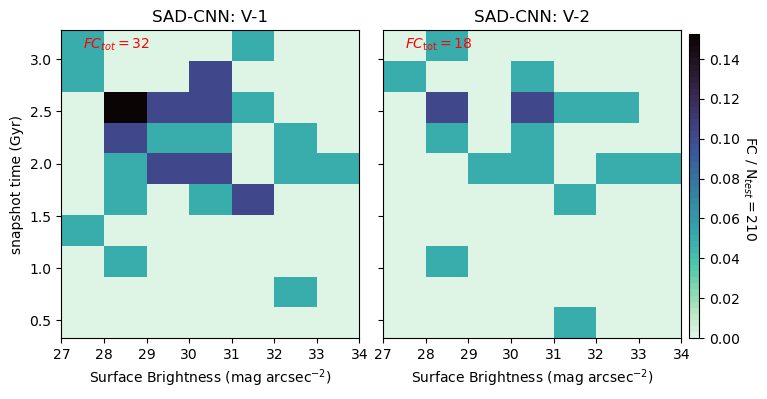}
    \caption{Density distribution of all false cases. We present density for CNN: V-1 and V-2 in the panels, respectively. The false cases are analyzed as a function of surface brightness and snapshot lookback time. This arrangement allows for a detailed examination of the interplay between these parameters in the context of false cases. The total number of false cases ($FC_{tot}$) is 32 and 18, respectively. This is normalised with the number of false cases and the number of test sample (210 images)}.
    \label{fig:dens_all_proop}
\end{figure}

We now explore the properties of the images that have been wrongly classified. Our goal is to compare the results obtained from the two versions of the SAD CNN to improve the model. Figure \ref{fig:dens_all_proop} shows the distribution of false cases in the surface brightness vs. the snapshot time space. The left panel shows that false cases in the CNN V-1 version tend to be distributed about earlier snapshot lookback time ($> 1.5$ Gyr), and mainly between $28 >\mu^{lim}> 31$ mag arcsec$^{-2}$. In other words, even though the V-1 version provides a reasonably good performance with the previously unseen data set,  most false cases are distributed at higher snapshot times. The right panel shows the distribution of false cases obtained with the V-2 version. We can clearly see that, after training CNN with images at higher snapshot times, the performance improves significantly. This only highlights the fact that the training set used in the V-1 version was not sufficiently diverse.   

\begin{figure}
    \centering
    \includegraphics[width=1\linewidth]{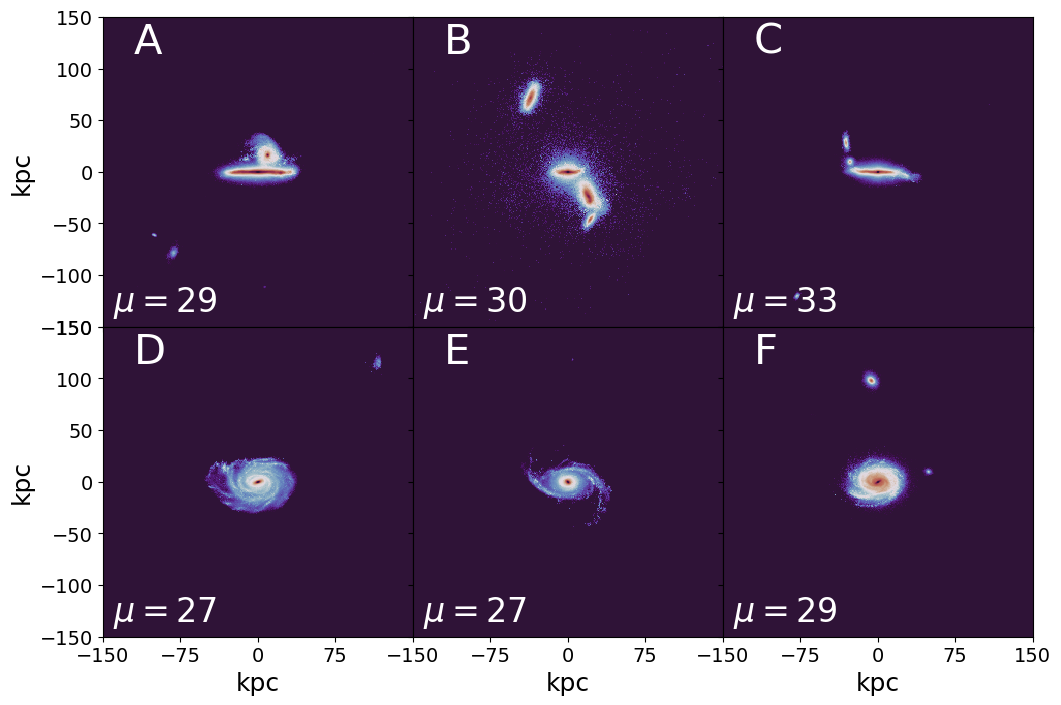}
    \caption{Examples of FP cases in SAD-CNN. The first row illustrates various luminous structures such as satellite accretion or merging satellites, that are misclassified as LSBF taken from different Auriga halos at different snapshot lookback times. The second row displays FP for halos at 45$^{\circ}$ inclination, these models present prominent disc features, such as spiral arms, which were mistakenly classified as stellar streams. All models present the $\mu_{lim}^{r}$ in mag arcsec$^{-2}$ of different Auriga Halos. The box has a side of 300 kpc.}
    \label{fig:example_sat}
\end{figure}

In Figure \ref{fig:example_sat}, we show examples of these misclassified galaxies, in particular we show the examples of FP cases. Each image has a length of 300 kpc by side. The edge-on halos shown have discs of different sizes. Interestingly, most of these examples show bright and still bound satellites with a signature of early tidal disruption. These particular configurations have proved difficult to correctly classify by our CNN. In the case of image B, we additionally observe a multiples satellites interacting with the host galaxy. Even though the larger number of images and the TL process applied to V-2 version has helped mitigate this effect, as we explore in the following section, the situation worsens when considering galactic discs with different inclinations.

\subsection{Galaxies with Inclinations}

\label{sec:4.3}
The edge-on projection of disk galaxies represents the best configuration for detecting the stellar streams using surface brightness maps. However, galaxies in the Universe are randomly oriented. As shown in the previous section, structure arising from the stellar disc can be misidentified as debris from disrupting satellites. A similar situation takes place when prominent spiral arms arise in the images due to the disc inclination. To explore whether such inclined configurations significantly reduce the accuracy of our CNN, which was exclusively trained with models in edge-on configurations, we consider the images generated for Group III, described in section \ref{training}. In this configuration, the mock images predominantly feature well-defined spiral structures, making them distinct from the earlier training sets.

\begin{figure*}[h]
    \centering
    \includegraphics[scale=0.45]{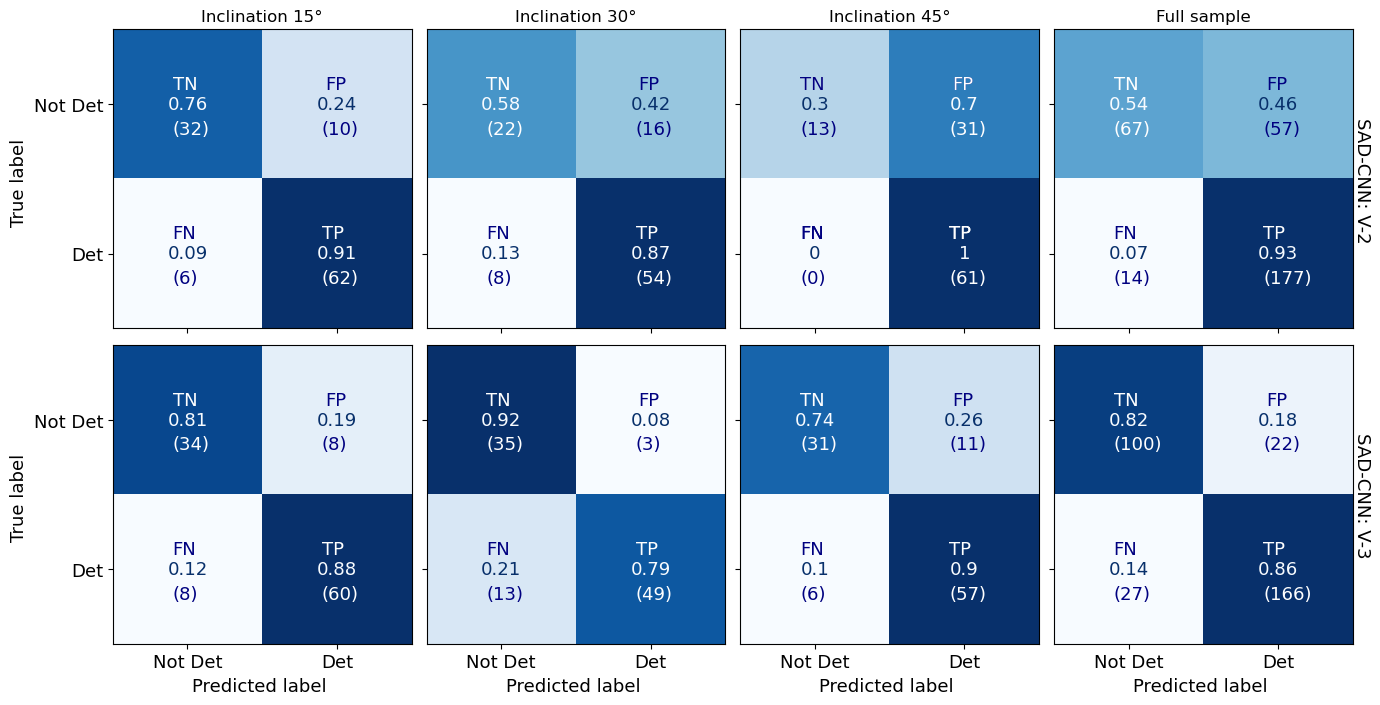}
    \caption{The confusion matrices shows the results before and after the TL process of predictions of stellar streams in galaxies at different inclinations. In the first and second row respectively. The rows indicate the degree inclination of the galaxies in Group III, the last row presents all samples. The rows inside the matrix represent the true labels of the galaxies, meanwhile, the columns represent the predicted labels of the model. The matrix right in both cases presents the complete sample with all inclinations.}
    \label{fig:matrix_incl}
\end{figure*}

In this new step we incorporate 5,264 new images using inclinations and rotations, applied to the initial sample with 1050 images at $z=0$. As mentioned previously, the authors made a visual inspection to identify faint features. The results are summarised in the confusion matrix, shown in Figure \ref{fig:matrix_incl}. Here, the top panels show the result obtained from  the SAD-CNN V-2, whereas the second row shows the results after applying TL. We will refer to this new model as SAD-CNN V-3. The first three columns show the results obtained for images of where galactic discs are inclined by 15, 30, and 45 degrees, respectively. The last one shows the results for the complete sample. As expected, the top panel clearly shows that the precision values of our V-2 model rapidly declines as we increase the inclination of the sample. The increase in FP cases is significant, jumping from 24 percent in the $i=15$ degrees sample to 70 percent in the $i=45$ degrees sample. The overall results are shown in the upper right panel. Even though the recall remains at an average of 0.74 for all cases, the set of images with detected LSBF by the CNN V-2 would be strongly contaminated by FP cases. 

The results obtained after applying the TL, presented in the bottom panels of Figure \ref{fig:matrix_incl}, show a significant improvement. Indeed, while we keep a high true positive detection rate, the fraction of false positives falls significantly at all inclinations. For example, at inclinations of 45 degrees, the FP rate shifts from 0.7 to 0.26 in the V-3 version. The overall results, shown on the bottom right panel, indicate TN and TP rate values of 0.82 and 0.86 percent, respectively. With respect to the previous average recall values, of 0.73, obtained with the V-2 version of the CNN, the new average increase at a value of 0.84.

\subsubsection{Most Common False Cases}

\begin{figure}[h]
    \centering
    \includegraphics[width=1\linewidth]{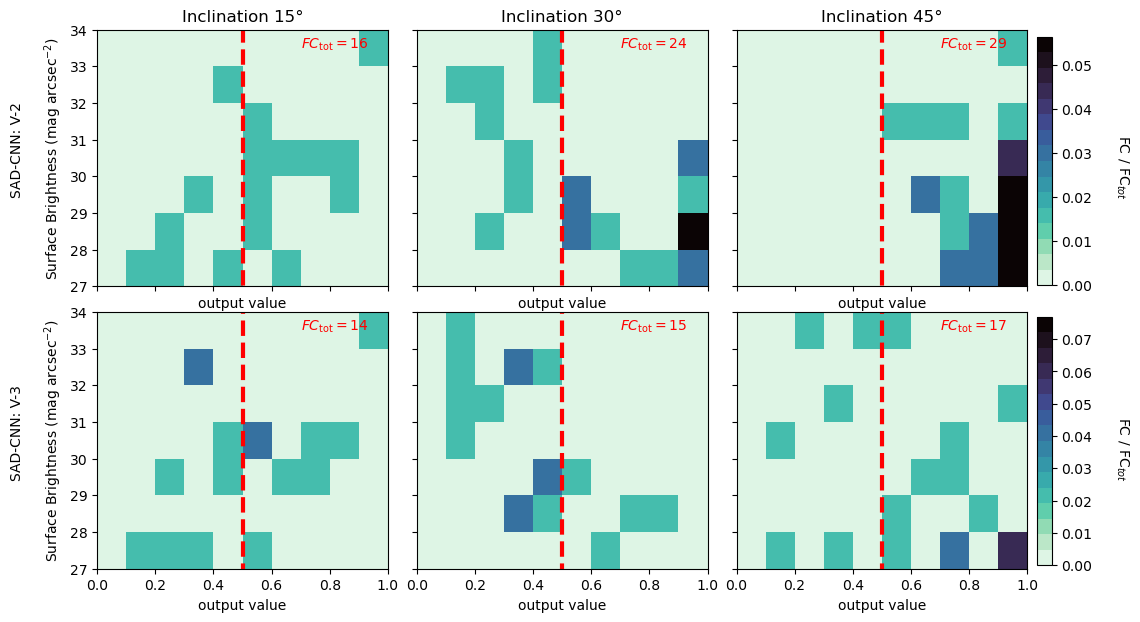}
    \caption{ Density Distribution of all false cases in the prediction of SAD-CNN. The panels present the distribution of the output value and the surface brightness limit in the r-band for the different inclinations. The red lines separate the not-detection zone (threshold $<0.5$) and detection zone (threshold$>0.5$). All panels have been normalized by the total number of images in the test set}.
    \label{fig:dens_incl}
\end{figure}

As previously done, we now explore the main characteristics of images associated with false cases. In Figure \ref{fig:dens_incl} we present the distribution of false cases at different surface brightness values as a function of the CNN output value. The results from the V-2 and V-3 models are shown in the first and second rows, respectively. The three columns indicate the distributions obtained at inclination angles of 15$^{\circ}$, 30$^{\circ}$, and 45$^{\circ}$, respectively. The red dashed line shows the threshold value used to select positive and negative detections. As a result, the left half of these panels are associated with FN cases, whereas the right side with FP cases. All panels have been normalized by the total number of images in the test set. At low inclinations (15 degrees) a comparison between the V-2 and V-3 reveals no significant differences. The false cases are nearly uniformly distributed in SB limiting magnitude. However, as we increase the inclination the differences become significant. We can clearly see that the CNN V-2 model (top panels) starts to produce false positive detections with high output model values. This is particularly true for $27 \geq \mu_{\rm lim} < 31$ mag arcsec$^{-2}$. In other words, the model assigns high confidence scores to images that do not actually have LSBF. In Figure \ref{fig:example_sat} we show examples of images that resulted in high confidence values as false positive detections by the CNN V-2 model. Notice that, as expected, all images present noticeable spiral arm patterns that are interpreted by the CNN  as stellar streams. This limitation is subsided after applying TL and, thus, further training the models with inclined images. The bottom panels of Fig.~\ref{fig:dens_incl} reveal that this overdensity of false positive cases, especially notorious at inclinations of 45 degrees in the V-2 version, is not present any longer. 

% FIGURA HORIZONTAL
%FIGURA VERTICAL (1 COL)
%\begin{figure}
%    \centering
%    \includegraphics[width=1\linewidth]{Fig/density_incl2.png}
%    \caption{ Density Distribution of all false cases in the prediction of SAD-CNN. The panels present the distribution of the output value and the surface brightness limit in the r-band for the different inclinations. The red lines separate the not-detection zone (threshold $<0.5$) and detection zone (threshold$>0.5$). The values are normalised according to the total false cases in each case.}
%    \label{fig:dens_incl}
%\end{figure}

\section{SAD-CNN Application}
\label{sec:Results}

In order to test the accuracy of the final model, i.e. CNN V-3, we created two new data sets comprising images that had not been analysed during the previous training process, at any stage. In particular, we considered input data images of the Auriga models at a snapshots that had not previously been used in Group II (see Sec. \ref{groups}). The redshifts for these tests is $z = [0.06, 0.09, 0.11, 0.14, 0.17, 0.2, 0.23, 0.26, 0.29, 0.33, 0.36]$. The first set consists of images with the galactic disk oriented edge-on. The galaxies were randomly rotated about their symmetry axis. This set comprises 1246 images never used in the learning process. As a second and more stringent test, we generated a set of stellar discs images at $z=0$ with randomly inclined between $0$ to $45$ degrees, following a uniform distribution for the inclination angle. This last step is necessary as we count with first group at different inclinations, without applying random rotations, the orientation of the galaxies at different snapshots in the cosmological box reference frame would be correlated. As a result, we produce a new test set comprising a total of 1880 mock images, randomly oriented and at different times. This exercise allows us to quantify the performance of our different CNN models in a more realistic scenario.

\subsection{Application in edge-on halos}

\begin{figure}[h]
    \centering
    \includegraphics[width=1\linewidth]{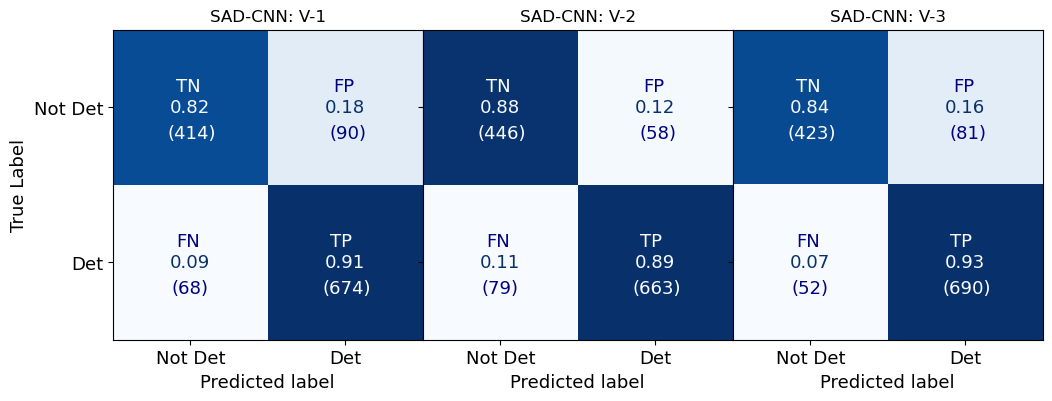}
    \caption{The confusion matrices of the results of the different versions of the SAD-CNN. The tree panels present each version, respectively. The rows inside the matrix represent the true labels of the galaxies. Meanwhile, the columns represent the predicted labels of the model.}
    \label{fig:edgeon_test}
\end{figure}

This test focuses on images where galactic discs are oriented edge-on. We evaluate the three versions of the SAD-CNN and compare their performance. The results of this test are summarised in Figure \ref{fig:edgeon_test}, where we show the resulting confusion matrices. The left, middle and right panels show the performance of three versions of the SAD-CNN: V-1, V-2, and V-3, respectively. As expected for this image test set, all versions of the SAD-CNN exhibit consistently high accuracy. The results were similar for all three versions of the model. They excel in identifying LSBF in our mock images, with a true positive rate of $\sim 90\%$.  On the other hand, the false positive cases reach a $16\%$ of the images that were classified as non-detection by our visual classification. As previously discussed, typically these cases are associated with structures arising from the stellar disc, or elongated satellites that can be observed in the mock images. The statistics of the three models provide an average precision, recall and F1-score of $\sim 0.88$.   

%%%%%%%%%%%%%%

\subsection{Application on inclined disc}

In our second test, we classify a set of images in which the galactic discs are randomly inclined. As before, we generated the images considering snapshots of the simulation that were not used during the training phase. 

\begin{figure}[h]
    \centering
    \includegraphics[width=0.48\textwidth]{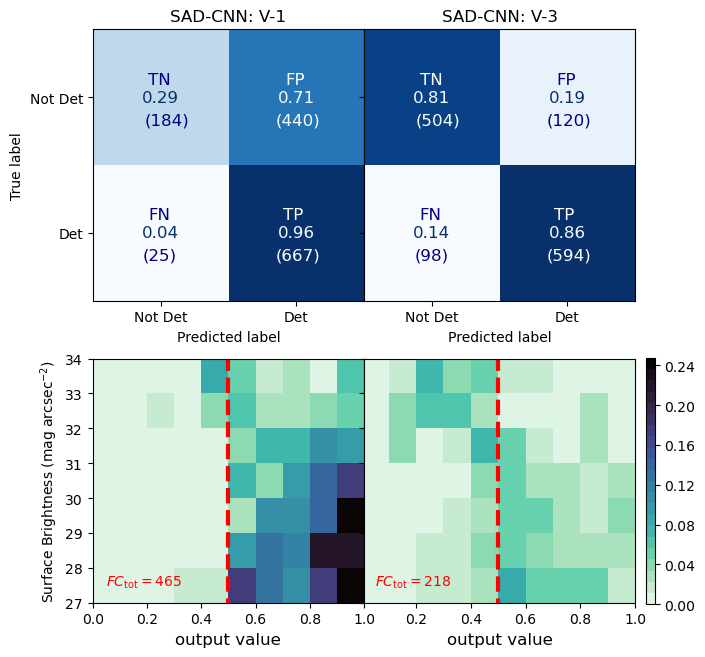}
    \caption{Top row: The confusion matrices of the results of versions 1 and 3 of the SAD-CNN. The columns present each version. The rows inside the matrix represent the true labels of the galaxies, meanwhile, the columns represent the predicted labels of the model. 
    Bottom row: Examples of the false cases for the different versions over the sample with models of galaxies in edge-on projection never seen before. These density plots are normalised with the total false cases in each panel.}
    \label{fig:inclined_test}
\end{figure}

The top panels of Figure \ref{fig:inclined_test} show the confusion matrices obtained from this exercise. For clarity, here we only compare the results of the CNN V-1 and V-3 models. As expected, the V-1 version of the model can very efficiently identify images with LSB features, with a true positive rate of $96\%$ (top left panel).  However, it fails at classifying images with non-detections. Indeed, we obtain a false positive rate of $71\%$. This is mainly due to the inability of the V-1 version of the model to deal with stellar disc substructures such as strong spiral arms. The bottom left panel shows the density distribution of false cases as a function of SB limiting magnitude and the CNN output value. As seen in Sec. \ref{sec:4.3}, the highest density peaks of false cases are associated with high CNN output values and SB magnitudes that go from $\approx 27$ to  31 mag arcsec$^{-2}$. This indicates that this version of the CNN produces false positive cases with high certainty. The top right panel of Fig. \ref{fig:inclined_test} shows the confusion matrix obtained after applying the V-3 version of the model to the same test set. We can clearly appreciate a significant improvement with respect to the V-1 version. We maintain a very high true positive rate, of $86\%$, and reduced the false positive rate to $19\%$.The bottom right panels shows the distribution of false cases in the same space as before. Notice that false cases of high CNN output values no longer concentration. Instead, they now present a more uniform distribution. The improvement can also be seen in the values of the different metrics, listed in Table \ref{tab:tab_statistic7}. Notice how the average precision, recall and F1-score go from 0.4, 0.65 and 0.59 in the V-1 version, respectively, to values of 0.8, 0.83 and 0.82 in the V-3 version.

The results of this experiment not only demonstrate the accuracy of our CNN, but also highlight how important consider training data sets as diverse as possible.
These datasets should incorporate not only images with different faint substructure morphologies but also consider other LSBF associated with the host galaxy that could be confused by a CNN as debris from disrupted satellites 

\begin{table}
    \centering
    \resizebox{\columnwidth}{!}{
    \begin{tabular}{|c|l|c|c|c|c|}
    \hline
      SAD       &          & Precision & Recall & F1-score & support\\
         \hline
                & Not Det   & 0.92 & 0.35 & 0.51 & 1143 \\
      V-1   & Det      & 0.49 & 0.95 & 0.64 & 737 \\
                & average  & 0.70 & 0.65 & 0.57 & 1880 \\
                & accuracy & &           & 0.59 & 1880 \\
        \hline
                 & Not Det   & 0.92 & 0.63 & 0.74 & 1143 \\
      V-2   & Det      & 0.61 & 0.91 & 0.73 & 737 \\
                 & average  & 0.76 & 0.77 & 0.74 & 1880 \\
                 & accuracy & &           & 0.74 & 1880 \\
         \hline
                & Not Det   & 0.89 & 0.82 & 0.85 & 1143 \\
      V-3   & Det      & 0.75 & 0.84 & 0.79 & 737 \\
                & average  & 0.82 & 0.83 & 0.82 & 1880 \\
                & accuracy & &           & 0.82 & 1880 \\
        \hline
    \end{tabular}
    }
    \caption{The summary of the values in the metrics obtained for the dataset with a total of 1880 models at a random inclination, rotation, and snapshot time. The values obtained for each version of the convolutional neural network respectively.}
    \label{tab:tab_statistic7}
\end{table}

%%%%%%%%%%%%%%%%%%%%%%%%%%%%%%%%%%%%%%%%%%%%%%%%%%%%%%%%%%%%%%%%%
\section{Discussion}
\label{discusion}

The detection of low surface brightness features in galaxy halos, such as stellar streams, shells and plumes has attracted significant attention. These structures, remnants of the accretion of satellite galaxies, offer a unique window into a galaxy's merger history and the processes shaping its stellar halo. Historically, such structures were detected by visual inspection of carefully reduced images in different photometric bands. Works such as \citet{2010Marinez-Delgado, Mouhcine2010, 2013ApJ...765...28A, Monachesi2014, 2018A&A...614A.143M, 2019A&A...632A.122M, MartinLSST2022, Martínez-Delgado2023, Juan_MiroCarretero2023}, provided the first relatively large data sets of deep observations of stellar halos where stellar streams were possible to detect through visual inspection. Several numerical analysis followed \citep{bullock_johnston2005, 2018ApJ...862..114S, MartinLSST2022, Vera-casanova2022, 2024arXiv240903585M}, placing constraints and providing predictions with respect to the information that can be extracted from the brightest detectable features. With the advent of deeper and larger observational surveys, as well as more sophisticated numerical simulations, the field has increasingly turned to automated methods to analyse the growing datasets. Among these, convolutional neural networks (CNNs) have emerged as powerful tools for image analysis and pattern recognition, particularly in astronomy. The Stream Automatic Detection with Convolutional Neural Networks (SAD-CNN) framework presented here contributes to this growing body of work by offering a method specifically optimized for detecting LSBF. While SAD-CNN achieves high accuracy in identifying LSBF in simulated data, it is important to contextualize its performance within the broader landscape of similar approaches and the challenges inherent to such tasks.

A significant strength of SAD-CNN lies in its carefully designed training methodology. The model was progressively trained, starting with simplified scenarios (e.g., edge-on galaxies at $z=0$) and later incorporating more complex configurations, such as varying galaxy inclinations and redshifts. This iterative approach, coupled with TL, allowed the model to adapt to increasingly challenging datasets, achieving precision and recall metrics above $80\%$ in most scenarios. Additionally, the use of a well-balanced training dataset was critical for mitigating biases, ensuring that the CNN accurately distinguished between galaxies with and without streams, at the surface brightness boundaries that have the strongest detection rates.

Our results not only showed that the SAD-CNN approach is a viable avenue to efficiently and accurately detect LSB features on large surveys, but also clearly highlighted the relevance that the diversity and quality of the training data set remains critical. Misclassification, such as the false cases identification of spiral arms as streams, highlighted the importance of refining feature recognition in this type of model. Insufficient representation of complex morphologies or configurations can limit the power of the model. Extending the dataset to include interacting galaxies, noise effects, and other real-world complexities through additional TL steps would address some of these limitations and further improve the robustness of the model.

The application of machine learning to the detection of LSB features is not unique to SAD-CNN. While some efforts, such as \citet{Farias2020, Baxter2021, Fontirroig2024}, have employed CNNs for galaxy morphology classification or related tasks, these methods do not specifically address the detection of LSBF in stellar halos. Recent work by \citet{Gordon2024} has made significant progress in this area by developing a CNN-based framework to classify LSB tidal features into distinct categories, such as streams, shells, and diffuse debris. Their approach, applied to synthetic images with noise and realistic surface brightness thresholds, demonstrated the ability of CNNs to identify and classify tidal features with high accuracy. SAD-CNN builds on the progress made in these studies by focusing on improving classification through systematic training. In this way, these tools can be applied to the challenges of LSBF detection, including faint and diffuse stellar halos. A particularly comparison is with \citet{2021MNRAS.504..372B}, who used CNNs to identify post-merger galaxies in simulations. While their primary focus was on classifying post-merger systems, this work is closely related, as post-mergers often leave behind stellar streams.  Both studies underscore the utility of CNNs in handling complex astrophysical structures, but SAD-CNN extends this by explicitly focusing on the detection of LSBF and optimizing its methodology for varying surface brightness limits. Additionally, methods employing phase-space analysis, such as the work of \citet{Kamdar2021}, introduce complementary approaches for identifying stellar streams using kinematic and dynamical information. Their findings indicate that while stellar streams in the Galactic disk can provide valuable insights into the Milky Way's dynamical history. This suggests that while CNN-based detection methods like SAD-CNN focus on direct image classification, integrating kinematic constraints from phase-space studies could offer a more comprehensive understanding of the origin and evolution of faint substructures.

As upcoming large-scale surveys, such as those conducted by the Vera C. Rubin Observatory, begin to deliver unprecedented volumes of deep imaging data, the role of automated tools like SAD-CNN will become increasingly critical. These tools not only accelerate the detection process but also enable statistical analyses of stellar streams across diverse galaxy populations. However, ensuring the reliability of these methods when applied to observational data will require careful calibration and validation, particularly to account for observational artifacts, noise, and varying spatial resolutions. In addition, a diverse and large training data set will be required to further optimize the efficiency and accuracy. As a follow-up step, we plan to apply SAD-CNN to a much larger mock observational data set extracted from the IllustrisTNG-50 simulations.  

The combination of CNN with complementary techniques, such as Bayesian inference or clustering algorithms, could provide a more comprehensive framework for studying stellar streams. This integration would enable more robust estimates of physical parameters, such as stream age, mass, and orbital history, which are critical for constraining galaxy formation models.

In summary, SAD-CNN represents an important step forward in automating the detection of LSB features in galaxy halos. While its performance is promising, its effectiveness should be evaluated alongside other methods to fully appreciate its contributions and limitations. Future work should focus on diversifying training datasets, incorporating additional simulation and observational data, and exploring hybrid approaches to address the challenges of feature misclassification.

\begin{acknowledgements}
    AVC and FAG acknowledge support from Agencia Nacional de Investigación y Desarrollo (ANID) Fondo Nacional de Desarrollo Científico y Tecnológico (FONDECYT) Regular 1211370. FAG gratefully acknowledges support by the ANID BASAL project FB210003. D.P. acknowledges financial support from ANID through FONDECYT Postdoctrorado Project 3230379. RB is supported by the SNSF through the Ambizione Grant PZ00P2-223532. FvdV is supported by a Royal Society University Research Fellowship (URF R1 191703 and URF R 241005).
\end{acknowledgements}

\section*{DATA AVAILABILITY}

    The data underlying this article will be shared on reasonable request to the corresponding author.

\bibliographystyle{aa}
\bibliography{aa}

\onecolumn
\begin{appendix} %First appendix
\label{sec:app}
\section{Learning Curves}

The learning curves presented in the appendix illustrate the training progress of the SAD-CNN model by tracking key performance metrics over multiple epochs. These curves typically show how the loss function and accuracy (or other evaluation metrics) evolve for both the training and validation datasets.

By analyzing these curves, we assess whether early stopping is needed to prevent overfitting, if additional training epochs could improve performance, or if adjustments in hyperparameters (e.g., learning rate) are necessary.

\begin{figure*}[h]
\includegraphics[width=1\linewidth]{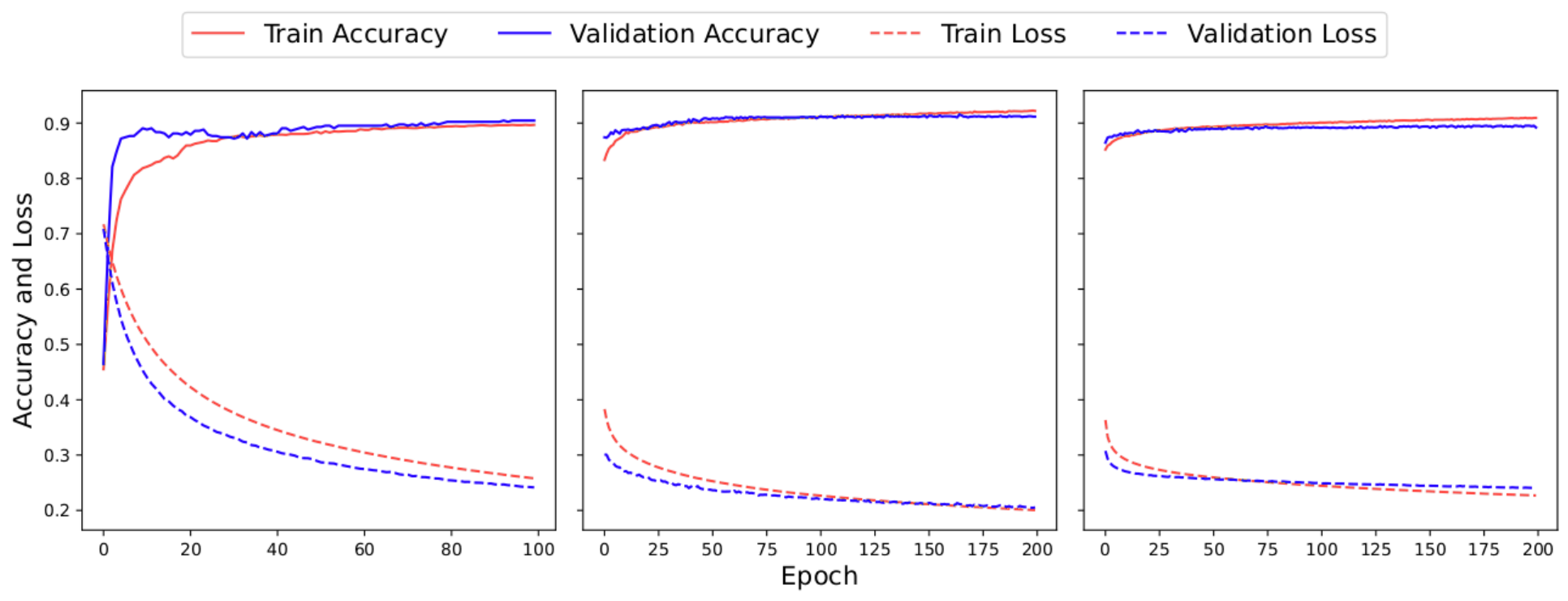}
\caption{Learning curves for the different versions of the SAD-CNN training process, panels correspond V-1, V-2, and V-3 respectively. The plot illustrates the model’s performance during training and validation for each network version. The x-axis represents the number of epochs, while the y-axis indicates the loss function values. The curves demonstrate how the model's error decreases over time for both the training (red lines) and validation (blue lines) sets. }
\end{figure*}

\end{appendix}

\end{document}